\documentclass[aps,prx,reprint,preprintnumbers,superscriptaddress,nofootinbib,longbibliography,floatfix]{revtex4-2}
\pdfoutput=1
\usepackage{rotating}
\usepackage{array}
\usepackage{amsmath}
\usepackage[normalem]{ulem}
\usepackage{slashed}
\usepackage{booktabs}
\usepackage[pdftex,table]{xcolor}
\usepackage{units}
\usepackage{xfrac}
\usepackage{mathtools}
\usepackage{empheq}
\usepackage[]{units}
\usepackage{multirow}
\usepackage{amssymb}
\usepackage{url}
\usepackage{comment}
\usepackage{physics}
\usepackage{color,soul}
\usepackage{bbm}
\usepackage{subcaption}
\usepackage{adjustbox}
\usepackage[T1]{fontenc}
\usepackage{enumerate}
\usepackage{booktabs}
\usepackage{multirow}
\usepackage{graphicx} 

\usepackage{hyperref}
\hypersetup{
  colorlinks=true,
  citecolor=blue,
  linkcolor=blue,
  urlcolor=blue
}

\begin{document}

\title{\textsc{OmniMol}: Transferring Particle Physics Knowledge\\to Molecular Dynamics with Point-Edge Transformers}

\author{Ibrahim Elsharkawy}
\email{ibrahim.elsharkawy@mail.utoronto.ca}
\affiliation{Department of Physics, University of Toronto and Vector Institute, Toronto, ON, Canada}
\affiliation{National Energy Research Scientific Computing Center (NERSC), Lawrence Berkeley National Laboratory, Berkeley, CA, USA}

\author{Vinicius Mikuni}
\email{vmikuni@hepl.phys.nagoya-u.ac.jp}
\affiliation{Nagoya University, Kobayashi-Maskawa Institute, Aichi 464-8602, Japan}

\author{Wahid Bhimji}
\email{wbhimji@lbl.gov}
\affiliation{National Energy Research Scientific Computing Center (NERSC), Lawrence Berkeley National Laboratory, Berkeley, CA, USA}

\author{Benjamin Nachman}
\email{nachman@stanford.edu}
\affiliation{Department of Particle Physics and Astrophysics, Stanford University, Stanford, CA 94305, USA}
\affiliation{Fundamental Physics Directorate, SLAC National Accelerator Laboratory, Menlo Park, CA 94025, USA}

\begin{abstract}
We present \textsc{OmniMol},  a state-of-the-art all-to-all transformer-based small molecule machine-learned interatomic potential (MLIP). \textsc{OmniMol} is built by adapting \textsc{Omnilearned}, a foundation model for particle jets found in high-energy physics (HEP) experiments such as at the Large Hadron Collider (LHC).  \textsc{Omnilearned} is built with a Point–Edge–Transformer (PET) and pre-trained using a diverse set of one billion particle jets.  It includes an interaction-matrix attention bias that injects pairwise sub-nuclear (HEP) or atomic (molecular-dynamics) physics directly into the transformer’s attention logits, steering the network toward physically meaningful neighborhoods without sacrificing expressivity. We demonstrate \textsc{OmniMol} using the oMol dataset and find excellent performance even with relatively few examples for fine-tuning. Further, due to architectural transfer from \textsc{Omnilearned}, we demonstrate uniquely fast inference. This study lays the foundation for building interdisciplinary connections given datasets represented as collections of point clouds.
\end{abstract}

\maketitle

\vspace{10mm}

\section{Introduction}
\label{sec:intro}

Machine learning (ML) has transformed how we represent, simulate, and analyze complex physical systems. Two domains where this transformation is particularly visible are particle physics~\cite{Radovic2018_MLHEP,Karagiorgi2022_MLNewPhysics} and molecular chemistry~\cite{vonLilienfeld2020_MLQM,Behler2021_HDNNP} (alongside related topics like molecular structure~\cite{Jumper2021_AlphaFold}).
The former pursues the fundamental structure of matter
while the latter is the basis for materials design and a number of other highly impactful applications.  Although vastly different in energy scale, both subjects are governed by quantum theories and share a common data structure: (variably-sized) sets of particles in (phase) space.  This is a type of structured \textit{point cloud}. Our hypothesis is that a foundation model built for point clouds in particle physics will be useful for ML tasks operating on point clouds in computational chemistry.  By construction, a foundation model for particle physics should improve diverse ML tasks across particle physics, but it is not known if transferability extends to completely different scientific disciplines.


To test the hypothesis, we adapt the recent \textsc{Omnilearned} particle physics foundation model~\cite{Mikuni:2024qsr,Mikuni:2025tar,Bhimji:2025isp} to molecular dynamics (MD).
\textsc{Omnilearned} is trained to classify and generate particle jets, which are a ubiquitous observable of high energy collisions. Physically, these jets emerge from the confining nature of Quantum Chromodynamics (QCD) acting on energetic quarks and gluons. Due to their complexity and relation to precision observables of the standard model, jets have been driving ML developments in particle physics~\cite{Larkoski:2017jix,Kasieczka:2019dbj,Feickert:2021ajf}.  We fine-tune \textsc{Omnilearned} on MD simulations to create \textsc{OmniMol} by exploring multiple approaches to fine tuning foundation models to cross domains.
%


Our goal is to develop a machine-learned interatomic potential (MLIP)~\cite{Smith2017_ANI1,Yao2018_TensorMol,Behler2007_BPNN,Schutt2017_SchNet,Gasteiger2021_GemNet,Batzner2022_NequiP,Batatia2022_MACE,Liao2024_EquiformerV2,
Qu2024_ScalableMLIP,Fu2025_eSEN}.  MLIPs aim to approximate potential energy surfaces and their gradients (i.e. forces) at a fraction of the computational cost of traditional methods, such as density functional theory (DFT)~\cite{Mardirossian2017ThirtyYearsDFT}. The goal is to enable large-scale and long-horizon MD. Given atomic types and coordinates $\{\vec r\in\mathbb{R}^3\}$, an MLIP predicts $\mathbf E(\{\vec r_i\})$, the energy of a given configuration of atoms and $\vec{F}_i(\{\vec r_i\})$, the forces on each atom due to the rest. These predictions can be used to `rollout' the motion of a molecule over time. Accuracy is necessary but not sufficient. MD utility depends on stability, energy conservation (enforced or approximate), and consistency  across molecular compositions.  \textsc{OmniMol} is trained on OMoL25~\cite{levine2025openmolecules2025omol25}, a large-scale molecular dataset, which provides atom positions in $\mathbb R^3$ as inputs and $\mathbf E(\{\vec r_i\})$ and $\vec{F}_i(\{\vec r_i\})$ in $\mathbb R^3$ as labels.  

In this paper, we develop a novel point-edge-transformer (PET) MLIP architecture, containing an interaction-matrix attention bias for chemically informed attention, along with equivariant and conservation constraints (unrelated to \cite{pozdnyakov2024smoothexactrotationalsymmetrization,bigi2026pushinglimitsunconstrainedmachinelearned} despite the similar name).  Using this architecture, we build an all-to-all transformer-based small molecule MLIP with near state-of-the-art inference speed and performance.  This is the first demonstration of cross-discipline transfer for scientific point-cloud foundation models.

While foundation models for natural language and images have found widespread use, there are fewer examples of cross-domain models specifically for scientific data.  A number of recent works have explored multiphysics foundation models for image representations of dynamical systems~\cite{mccabe2025walrus,herde2024poseidon,mccabe2024multiple,liu2024prosefd,wiesner2025gphyt}.  Within the discipline of fundamental physics, the \textsc{OmniCosmos} model~\cite{mikuni2025omnicosmostransferringparticlephysics} has a similar setup to \textsc{OmniMol}, transferring particle physics point cloud pre-training to cosmological point clouds.  

This paper is organized as follows.  Section Ref.~\ref{sec:model} briefly reviews the \textsc{Omnilearned} model and training protocol.  Next, Sec.~\ref{sec:omnimol} introduces \textsc{OmniMol}.  Equivariance and energy conservation are discussed in Sec.~\ref{sec:equi}.  Model training is detailed in Sec.~\ref{sec:pretraining}.  Results are presented in Sec.~\ref{sec:results} and the paper ends with conclusions and outlook in Sec.~\ref{sec:conclusions}.

\section{\textsc{OmniLearn}(\,ed\,)}
\label{sec:model}

\paragraph{Backbone and heads.}
\textsc{OmniLearned} is composed of a general core called the the point-edge transformer (PET) body and a number of task-specific heads.  The model couples \emph{local} attention over $k$-nearest neighbors with physics-inspired pairwise features and \emph{global} transformer blocks whose attention matrices are additively biased by the same interaction terms. Four learnable tokens summarize jet information. Two task heads share the PET body: a classifier and a generator trained with diffusion/flow-matching style objectives~\cite{bhimji2025omnilearned}.

\paragraph{Inputs}
Each jet is represented as an unordered variable-sized set of its constituent particles. Each particle is specified by 1) kinematic information relative to the overall jet momentum,
2) discrete Particle Identification Numbers (PIDs) that indicate the type of particle (such as proton or neutron) and 3) additional per particle information that is not always available, such as the reconstructed origin of the particle in the detector.

\paragraph{Embeddings}
Each jet constituent is embedded into a token space given by:
\begin{equation}
\begin{split}
        \vec x_{embed} = \vec x_{embed}^{kinematic}+\vec x_{embed}^{PID}+\\ +\vec x_{embed}^{time}+\vec x_{embed}^{add}+\vec x_{embed}^{local}\,.
\end{split}
\end{equation}
 Kinematic information is embedded to $\vec x_{embed}^{kinematic}$ via a shared multilayer perceptron (MLP), and PID information $\vec x_{embed}^{PID}$ is embedded via a learned lookup table. For generation, the diffusion/flow time variable is embedded via an MLP and injected through $\vec x_{embed}^{time}$. Finally, $\vec x_{embed}^{add}$ is found by embedding the additional information via another MLP. 

\paragraph{Local attention.}
\begin{figure}
    \centering
    \includegraphics[width=\linewidth]{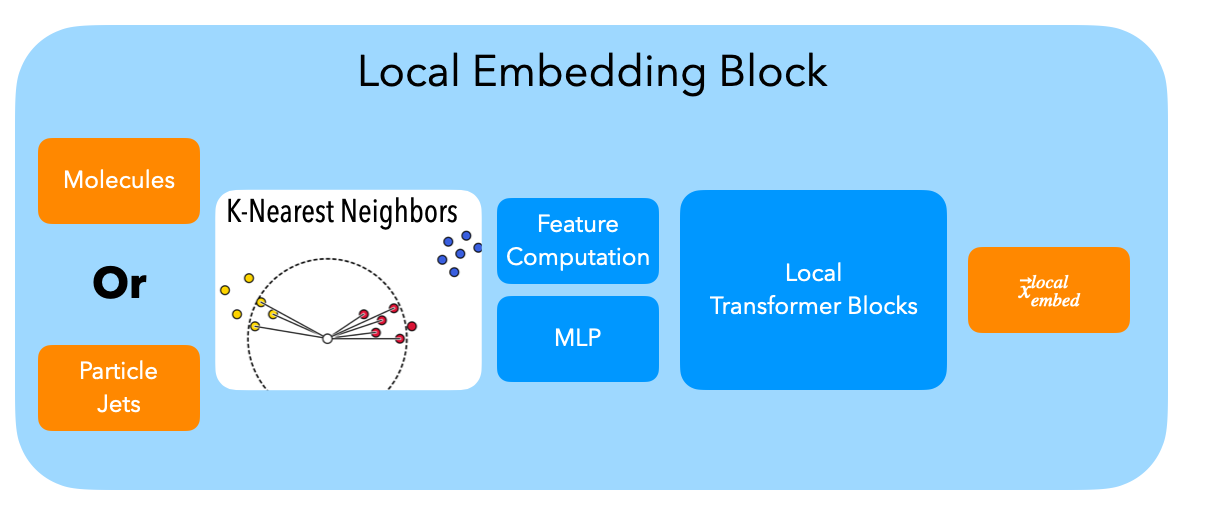}
    \caption{Local Embedding Block, where K-Nearest Neighbors for either Molecule or Jet point-clouds are computed along with pairwise features, these are fed to an MLP, and a small local transformer resulting in $\vec x_{embed}^{local}$}
    \label{fig:localembed}
\end{figure}
The last component of the embedding is $\vec x_{embed}^{local}$ and given by a local attention block. For each particle, 
%
a local neighborhood is built out of the K-nearest neighbors (with K${=}10$) using a Euclidean L2 notion of distance.
\textsc{Omnilearned} then builds a set of pairwise features $f(x_i,x_j)$  
followed by an MLP that transforms each of the K${}^2$ features $f(x_i,x_j)$ before being fed into a small local transformer block. This local block aggregates all pairwise information of the K-nearest neighbors for each jet constituent into a local embedding vector $\vec x_{embed}^{local}$. A diagram illustrating this block is found in Figure \ref{fig:localembed}.

\paragraph{Global attention and Bias}
After $\vec x_{embed}$ is computed for each jet constituent, the unordered set of $\vec x_{embed}$ is fed into large global all-to-all transformer blocks. These global self-attention components add an explicit bias to the attention logits found by computing $f(x_i,x_j)$ for all \textit{pairs} of particles, and embed via an MLP into $f(x_i,x_j)\rightarrow B_{ij}$, such that the logits are set to $A_{ij}^{*} \;=\; A_{ij} \;+\, B_{ij}$ for $B_{ij} \in \mathbb{R}$.
\paragraph{Output Heads}
\textsc{OmniLearned} is built with two output heads attached to the transformer body, a classifier head, and a generator head. The generator head is composed of two smaller all-to-all transformer blocks followed by a shared MLPs applied to each token. This head is by construction \textit{permutation equivariant}. The classifier head is composed of two "Token" Attention Blocks, which summarize jet information from the incoming body output into $n=4$ trainable tokens. These summary tokens are then flattened and fed into an output MLP, which collapses the dimensionality to the desired number of class labels.  

\paragraph{Multi-task objective.}
Pretraining optimizes a joint loss that combines (i) cross-entropy for supervised jet-type classification, (ii) a velocity-matching $\ell_2$ loss (flow matching, equivalent to diffusion) for generation, and (iii) a classification on noised inputs term to couple the two representations. To exploit unlabeled data, the classifier head is split to predict either jet type (when labels exist) or \emph{sample identity} (dataset classification) with equal weighting in the loss.

\paragraph{Pre-training dataset and Compute.}
\textsc{OmniLearned} is pretrained on \emph{over one billion} jets drawn from a unified suite of open datasets. Training runs on 32–512 A100 GPUs using the Perlmutter supercomputer, employs a global batch size of 4096 for three full passes over the 1B-jet corpus using the Lion optimizer~\cite{chen2023symbolicdiscoveryoptimizationalgorithms}. Released model sizes are small, medium, and large, with roughly 3M, 58M, 460M trainable weights.

\begin{figure*}
    \centering
    \includegraphics[width=.7\linewidth]{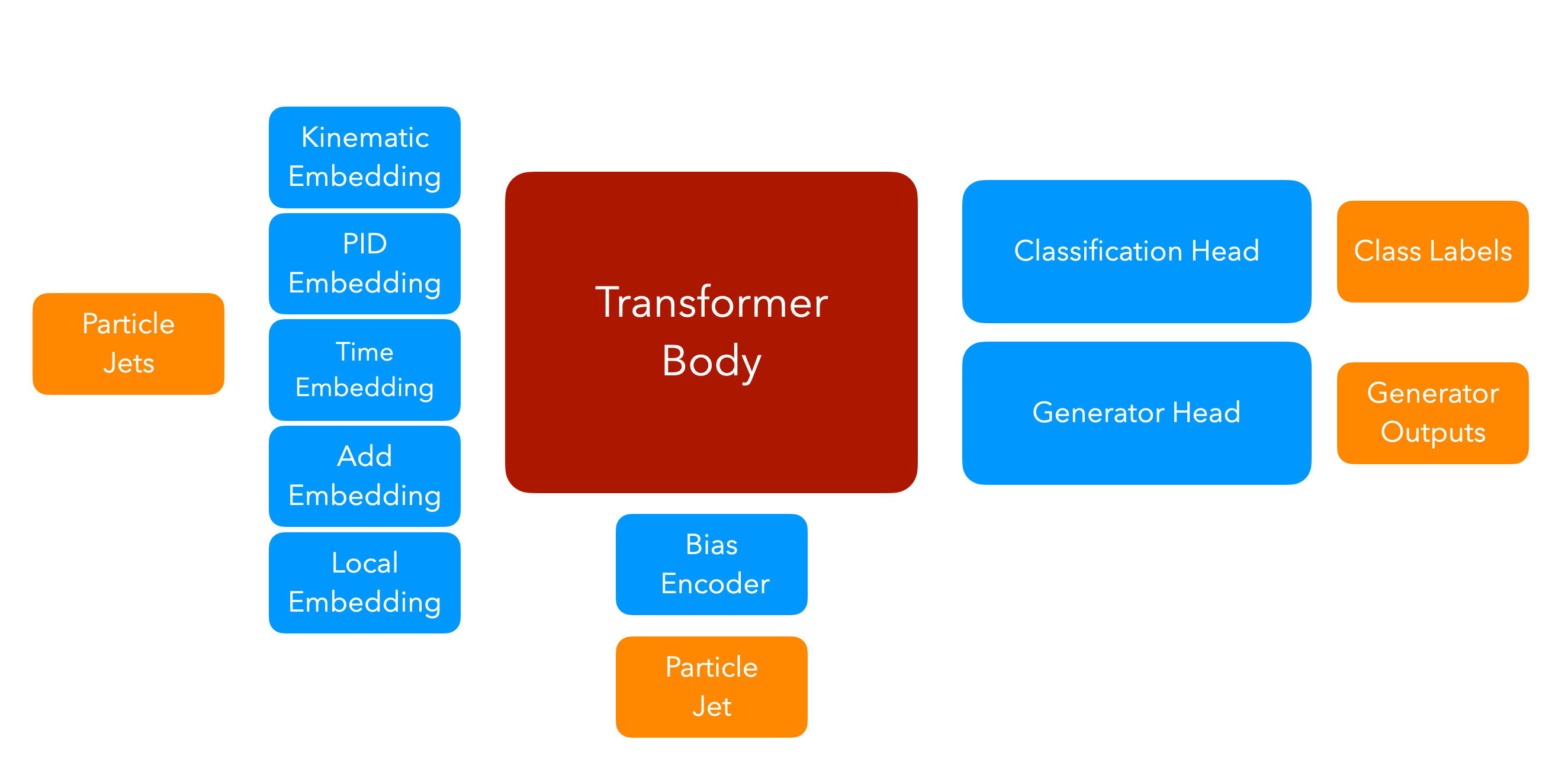}
    \caption{\textsc{OmniLearned}'s Architecture }
    \label{fig:OmniLearned}
\end{figure*}

More details regarding \textsc{OmniLearned}'s architecture and training can be found in the paper Ref.~\citep{bhimji2025omnilearned}. A graphical depiction of \textsc{OmniLearned}'s architecture can be found in Figure \ref{fig:OmniLearned}.
\section{OmniMol}
\label{sec:omnimol}
\begin{figure*}
    \centering
    \includegraphics[width=.65\linewidth]{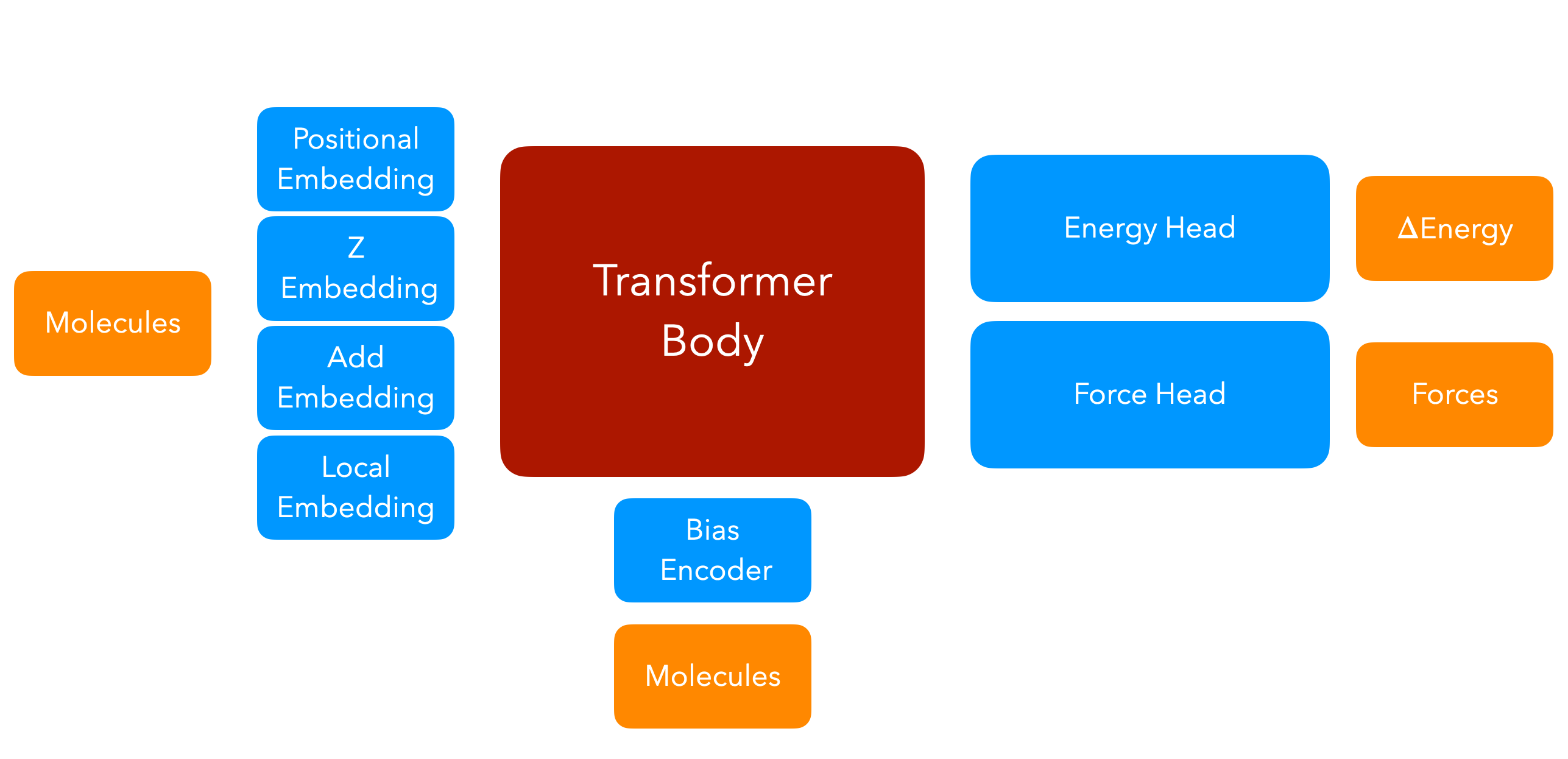}
    \caption{\textsc{OmniMol}'s Architecture}
    \label{fig:OmniMol}
\end{figure*}
We take \textsc{OmniLearned} and swap out the input encoders and output heads to ones that match a molecular prior. These changes define \textsc{OmniMol}. As we describe \textsc{OmniMol}, the similarities between the two problems become explicit.
\paragraph{Shared Backbone and Heads.}
\textsc{OmniMol} uses a similar input encoders, similar heads, and the same body as \textsc{OmniLearned}. The generative head, which is by construction permutation equivariant, is repurposed for per-atom direct force prediction. A copy of this same generative head is repurposed for energy prediction, outputting a per-atom energy correction $e_i$ that are summed over each molecule to find $\widetilde{\mathbf{E}}^* = \sum_i e_i$. The quantity $\widetilde{\mathbf{E}}^*$ can then be compared to the standardized molecule level label $\widetilde{\mathbf{E}}$, preserving an extensive prior.

\paragraph{Inputs}
The model is trained on oMol, which gives molecules as an unordered variable-sized set of atoms. Each atom contains 1) positional information given by Euclidean coordinates $\vec r\in\mathbb R^3$ relative to some origin 2) discrete Atomic Numbers ($Z$) and 3) additional per-atom charge and spin information \citep{levine2025openmolecules2025omol25}.

\paragraph{Embeddings}
Each atom is embedded into a token space given by:
\begin{equation}
\begin{split}
        \vec x_{embed} = \vec x_{embed}^{pos}+\vec x_{embed}^{Z} +\vec x_{embed}^{add}+\vec x_{embed}^{local}\,.
\end{split}
\end{equation}
 Positional information $\vec r_i$ is embedded to $\vec x_{i,embed}^{pos}$ via a shared MLP. Atomic Number $Z$ is embedded to $\vec x_{embed}^{Z}$ via a learned lookup table, and $\vec x_{embed}^{add}$ is found by embedding additional atom charge and spin via another MLP when available. 

\paragraph{Local attention}
The last component of the embedding is $\vec x_{embed}^{local}$ and given by a local attention block. For each atom, local neighborhoods defined by K-nearest neighbors (with K${=}15$) with respect to the atoms position vector are built, and \textsc{OmniMol} then computes pairwise physical features for each pair of atoms in the neighborhood,
\begin{equation}
\begin{split}
        f(\vec r_i,\vec r_j,\vec x_{i,embed}^{Z,add},\vec x_{j,embed}^{Z,add})=\bigg[\vec r_i-\vec r_j,  ~||\vec r_i-\vec r_j||,~\\ \frac{1}{||\vec r_i-\vec r_j||}, \frac{1}{||\vec r_i-\vec r_j||^2},~ \frac{1}{||\vec r_i-\vec r_j||^6},\\~\frac{g(\vec x_{i,embed}^{Z,add},\vec x_{j,embed}^{Z,add})}{||\vec r_i-\vec r_j||},\frac{g(\vec x_{i,embed}^{Z,add},\vec x_{j,embed}^{Z,add})}{||\vec r_i-\vec r_j||^2} ,\\\frac{g(\vec x_{i,embed}^{Z,add},\vec x_{j,embed}^{Z,add})}{||\vec r_i-\vec r_j||^6},
        \{RBF(\vec r_i,\vec r_j)_m|~\forall~ m\}
        \bigg],
\end{split}
\end{equation}
followed by an MLP that transforms each $f(\vec r_i,\vec r_j,\vec x_{i,embed}^{Z,add},\vec x_{j,embed}^{Z,add})$ before being fed into a small local transformer block. The quantity $g(\vec x_{i,embed}^{Z,add},\vec x_{j,embed}^{Z,add})$ is a learned function of atomic number embeddings and additional feature embeddings computed by an MLP. Lastly, $\{RBF(\vec r_i,\vec r_j)_k|~\forall~ k\}$ is a set of radial basis functions (RBFs) used to give the MLP maximal flexibility in learning important 2-point correlations. \textsc{OmniMol} uses Gaussian RBFs given by, 
\begin{equation}
    RBF(\vec r_i,\vec r_j)_m = \exp(-\frac{||\vec r_i-\vec r_j||^2}{2\sigma_m^2})\,,
\end{equation}
where $m$ ranges from $0$ to $30$, and $\sigma_m$ is evenly spaced on a range from $(0.1,10)$.
This local block aggregates all pairwise information of the K-nearest neighbors for each atom into a local embedding vector $\vec x_{embed}^{local}$. A diagram illustrating this block can again be found in Figure \ref{fig:localembed}.

\paragraph{Global attention and Bias}
After $\vec x_{embed}$ is computed for each atom, the unordered set of $\vec x_{embed}$ is fed into \textsc{Omnilearned}'s large all-to-all transformer blocks. These global self-attention blocks, once again, add an explicit bias to the attention logits found by computing the same chemical priors $f(\vec r_i,\vec r_j,\vec x_{i,embed}^{Z,add},\vec x_{j,embed}^{Z,add})$ for all \textit{pairs} of particles, and embed via an MLP into biases $f(\vec r_i,\vec r_j,\vec x_{i,embed}^{Z,add},\vec x_{j,embed}^{Z,add})\rightarrow B_{ij}$, such that the logits are again set to $A_{ij}^{*} \;=\; A_{ij} \;+\, B_{ij}$ for $B_{ij} \in \mathbb{R}$.  An illustration of \textsc{OmniMol}'s architecture can be found in Figure \ref{fig:OmniMol}. 
\section{Building in Invariance and Conservation}
\label{sec:equi}
\begin{figure*}
    \centering
    \includegraphics[width=.65\linewidth]{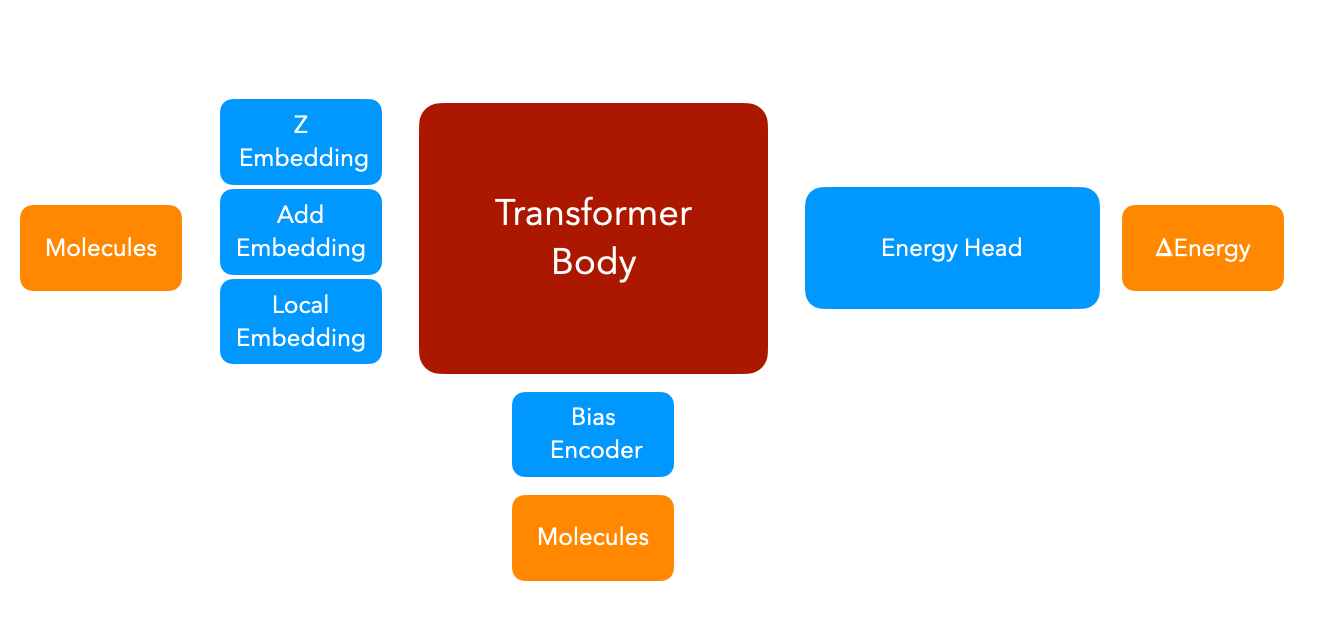}
    \caption{Conservative and Equivariant \textsc{OmniMol}'s Architecture}
    \label{fig:OmniMol_cons}
\end{figure*}
It is often desired to restrict a model with certain physical constraints/priors. For MLIPs, two constraints are desired: rotational equivariance (or invariance) and energy conservation.

Energy conservation implies that the forces on atom $i$ is given by:
\begin{equation}
    \vec{F}_i = \frac{dE(\{\vec r_j\})}{dr_i}\,.\label{eq:cons}
\end{equation}
Energy conservation is implemented in \textsc{OmniMol} by dropping the direct force head and compute forces via Eq. \ref{eq:cons} with backpropagation. Due to the required double backpropagation, energy conservation slows down training and increases the required computation significantly. We thus only train \textsc{OmniMol} small this way, \textit{leaving larger models to future work. }

The second constraint a MLIP is expected to satisfy is rotational equivariance. Physically, any rotation (or change of coordinates) should have no effect on the magnitude of the force on each atom, nor on the magnitude of the scalar energy for the configuration. That is, if a structure's coordinates transform as,
\begin{equation}
    \vec x\rightarrow Q\vec x\,, \label{eq:equiv}
\end{equation}
for some rotation matrix $Q$, forces and energies should transform via,
\begin{equation}
    E(\{\vec r_i\})\rightarrow E(\{\vec r_i\}),\quad \vec{F}_i(\{\vec r_i\})\rightarrow Q\vec{F}_i(\{\vec r_i\}). \label{eq:e_and_f_transform}
\end{equation}
\textsc{OmniMol} as described is not equivariant or , due to the many locations where raw coordinates or coordinate differences are fed to MLPs, which are not constrained to satisfy Eq. \ref{eq:equiv}, and due to the \textit{direct} force head not constrained to transform forces according to Eq. \ref{eq:e_and_f_transform}. To remedy this, we first use conservative \textsc{OmniMol}, we then drop the input encoder, and drop the $\vec r_i-\vec r_j$ term in the pairwise features $f(\vec r_i,\vec r_j,\vec x_{i,embed}^{Z,add},\vec x_{j,embed}^{Z,add})$ in the local and interaction block. 

We find that the removal of $\vec r_i-\vec r_j$ reduces performance significantly, and we remedy this loss by introducing angular information. In the local block, for each atom, we take its K-nearest neighbors as before. We then take two vectors defined by
\begin{equation}
    \vec v_1 = \vec r_i-\vec r_j, \quad \vec v_2 = \vec r_k-\vec r_m\,,
\end{equation}
where $r_i$, $r_j$, $r_k$ and $r_m$ are atoms neighboring the center of interest. We then compute the cosine between $\vec v_1$ and $\vec v_2$ and append all ${\text{K}\choose 2}$ such angles to the features fed to the small local transformer. This addition improves performance significantly, while retaining equivariance. The architecture of the equivariant and conserving variant of \textsc{OmniMol} can be found in Fig. \ref{fig:OmniMol_cons}.

\section{Training/Fine-Tuning on oMoL}

\label{sec:pretraining}

\subsection{Input and Label Pre-Processing}
We treat each molecule in oMoL as a variable-size point cloud of atoms with associated coordinates and atomic numbers. Let a molecule be specified by $\{\mathbf{r}_i, Z_i\}_{i=1}^{N},$
where $\mathbf{r}_i \in \mathbb{R}^3$ are Cartesian coordinates and $Z_i$ are atomic numbers. To remove global translation and constrain the scale, we perform a per-molecule centering by subtract the center
\begin{equation}
    \bar{\mathbf{r}} = \frac{1}{N} \sum_{i=1}^{N} \mathbf{r}_i\,,
\end{equation}
from all coordinates, $\tilde{\mathbf{r}}_i = \mathbf{r}_i - \bar{\mathbf{r}}$. 
Energies are processed to remove a simple, additive baseline corresponding to the heat of formation (a ``bag-of-atoms'' reference). For each molecule, the DFT total energy $E_{\mathrm{DFT}}$ is decomposed as
\begin{equation}
E_{\mathrm{DFT}} = \sum_i^NE_{Z_i} + \Delta E,
\end{equation}
where $E_{Z_i}$ is the energy of an isolated atom with a given $Z_i$, $N$ is the total number of atoms in the molecule. We use $\Delta E$ as the primary target, and further apply a global standardization across the training set:
\begin{equation}
\tilde{E} = \frac{\Delta E - \mu_{\Delta E}}{\sigma_{\Delta E}},
\end{equation}
where $\mu_{\Delta E}$ and $\sigma_{\Delta E}$ are the mean and standard deviation of $\Delta E$ over training molecules. This yields dimensionless energy targets with approximately unit variance. Finally, we standardize the components of $\widetilde{\mathbf{F}}_i$ across the training set to have zero mean and unit variance per Cartesian direction, yielding force targets that are numerically comparable to the standardized energies.
\subsection{Training Objective, Optimizer and Scheduler}
We train with the sum of the mean absolute error (MAE) of energies and the per component pet atom MAE of forces weighted by $\lambda_E=1$ and $\lambda_F=10$, such that our loss function is given by:
\begin{equation}
    \mathcal{L} = \lambda_E |\widetilde{\mathbf{E}}-\widetilde{\mathbf{E}}^*| + \frac{\lambda_F}{3N} \sum_i^N\sum_j^3|\widetilde{\mathbf{F}}_{ij}-\widetilde{\mathbf{F}}_{ij}^*|\,. 
\end{equation}

For the optimizer, we find the standard \textsc{AdamW}~\cite{loshchilov2019decoupledweightdecayregularization} to be effective with learning rate $\eta=1e-3$ for \textsc{OmniMol-s}, $\eta=3e-4$ for \textsc{OmniMol-m}, and $\eta=1e-4$ for \textsc{OmniMol-l}. We train with 32/128/512 A100-40GB GPUs for the three model sizes.  We keep other optimizer parameters such as $\beta_1,\beta_2=.95, .99$ and weight decay $1e-4$ fixed for all runs.

Finally, we employ a \texttt{OneCycle} learning-rate policy for all trainings, which ramps the learning rate up and then down over the course of training. This schedule encourages rapid exploration early on, followed by a controlled annealing phase that reduces overfitting and improves convergence \citep{smith2018superconvergencefasttrainingneural}. We train all oMol-4M runs for 100 passes of the data, and all oMol-100M/oMol-140M runs with 15 passes of the data, and choose the checkpoint with the best validation loss.

We consider two fine-tuning strategies, as described in the next two subsections.

\subsection{LoRA Fine-Tuning}
\begin{figure*}[t]
    \centering
    \includegraphics[width=0.49\textwidth]{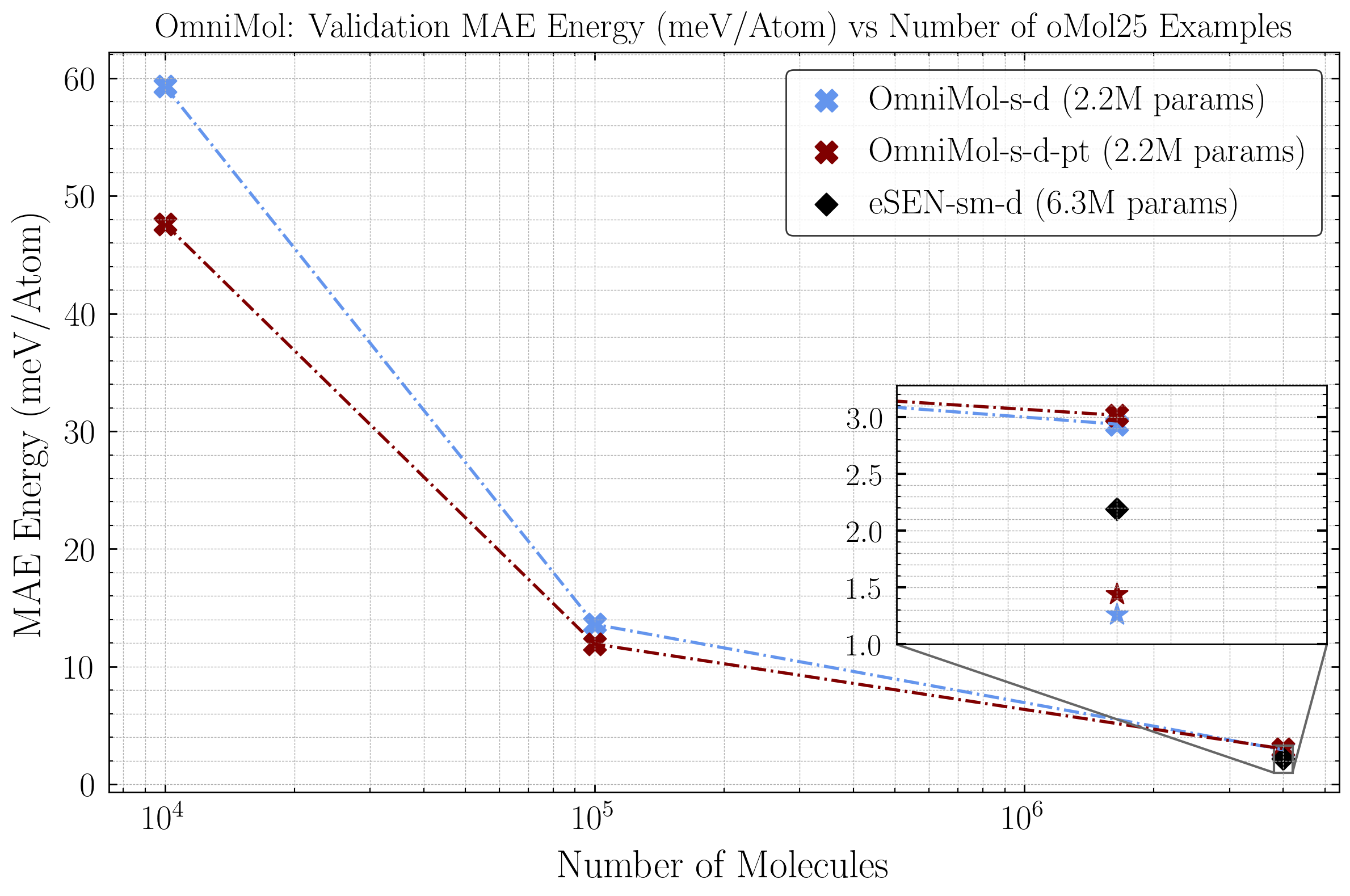}\hfill
    \includegraphics[width=0.49\textwidth]{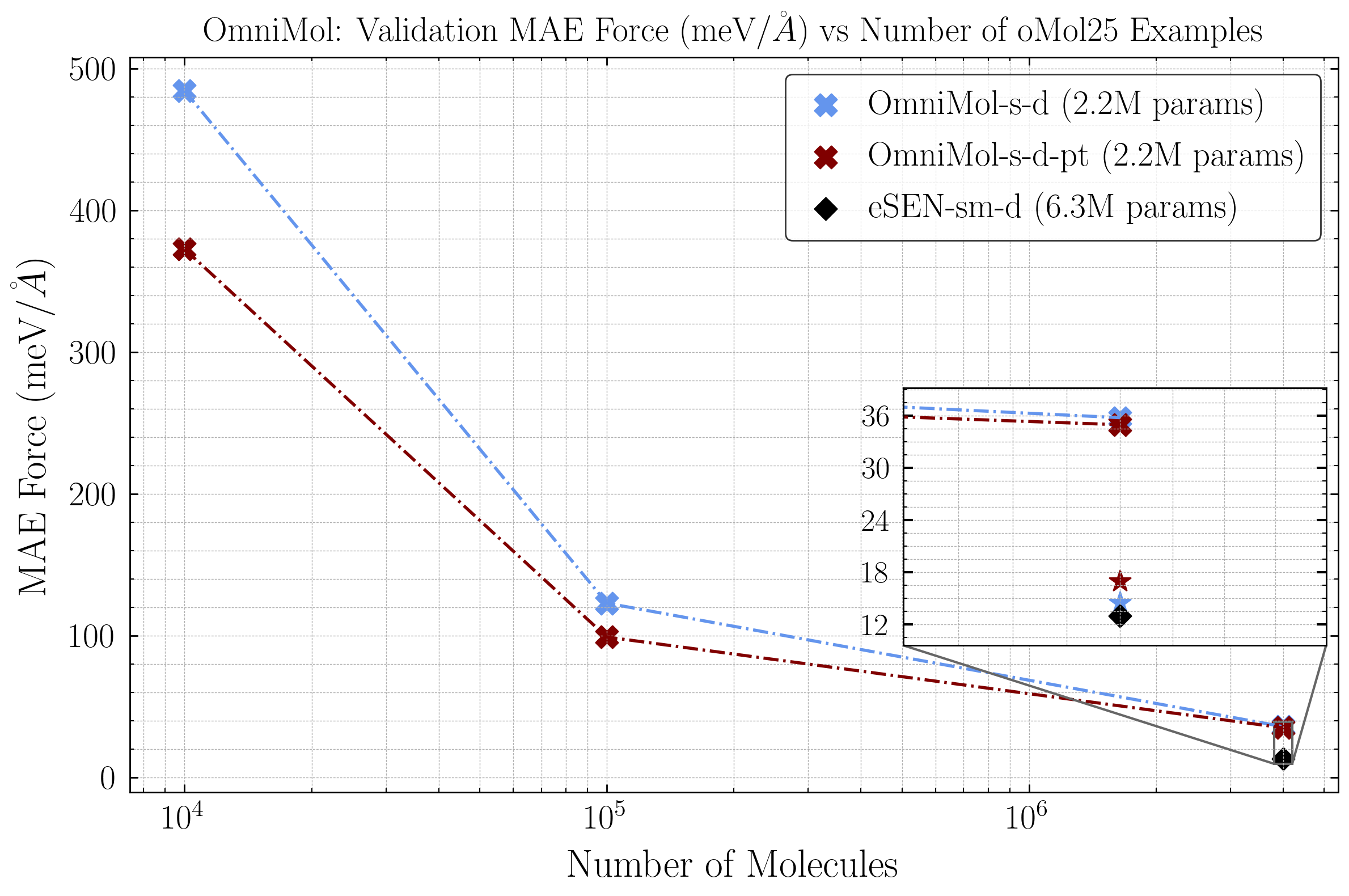}
    \caption{Scaling behavior for (left) energy and (right) forces of \textsc{OmniMol} direct small and medium pre-trained and from scratch. Finetuning with ten and one hundred thousand molecules on \textsc{OmniMol} small proceeds with LoRA, 4 million and \textsc{OmniMol} medium with full finetuning.}
    \label{fig:energy_scaling}
\end{figure*}
\begin{figure*}[t]
    \centering
    \includegraphics[width=0.49\textwidth]{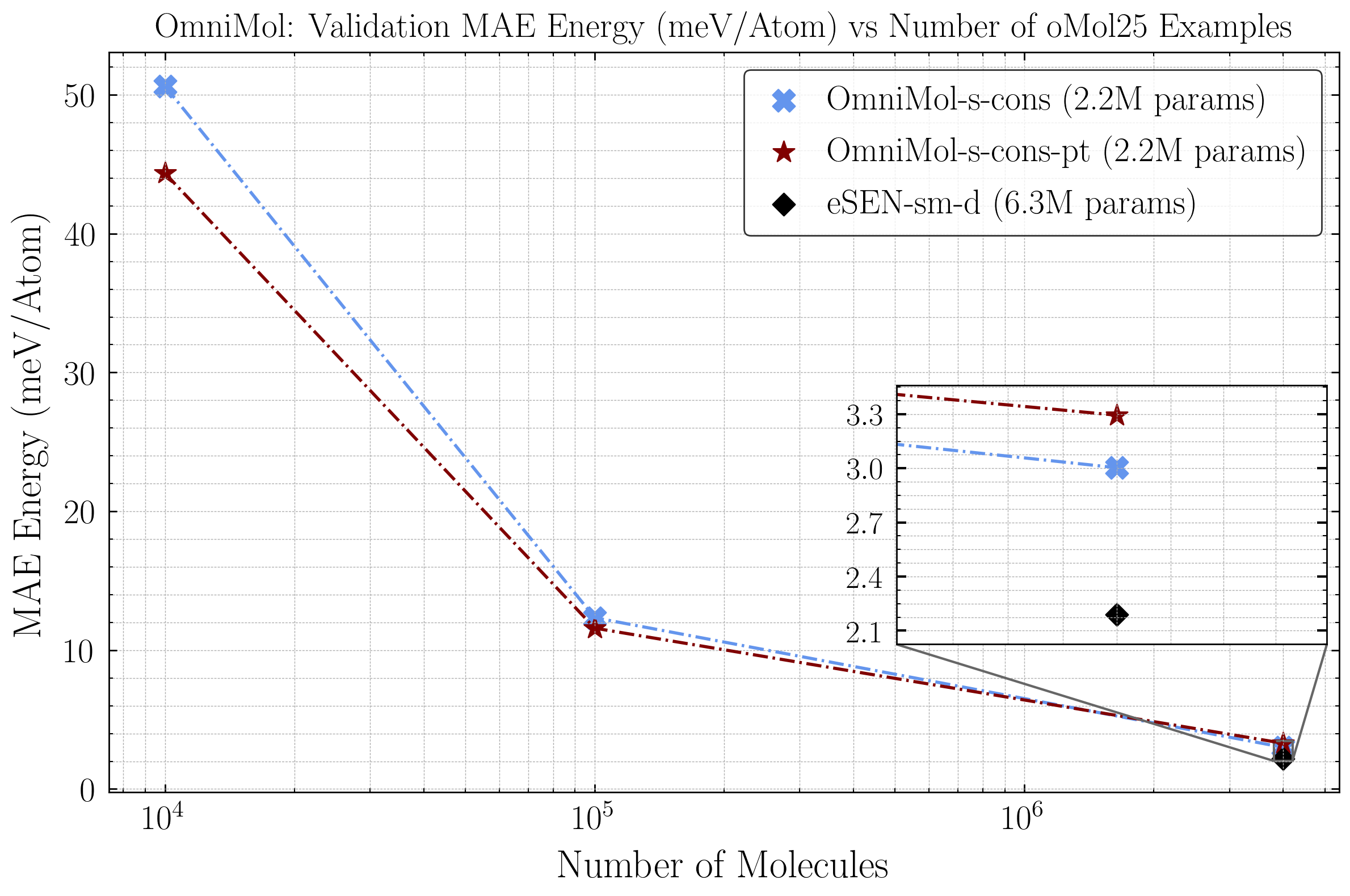}\hfill
    \includegraphics[width=0.49\textwidth]{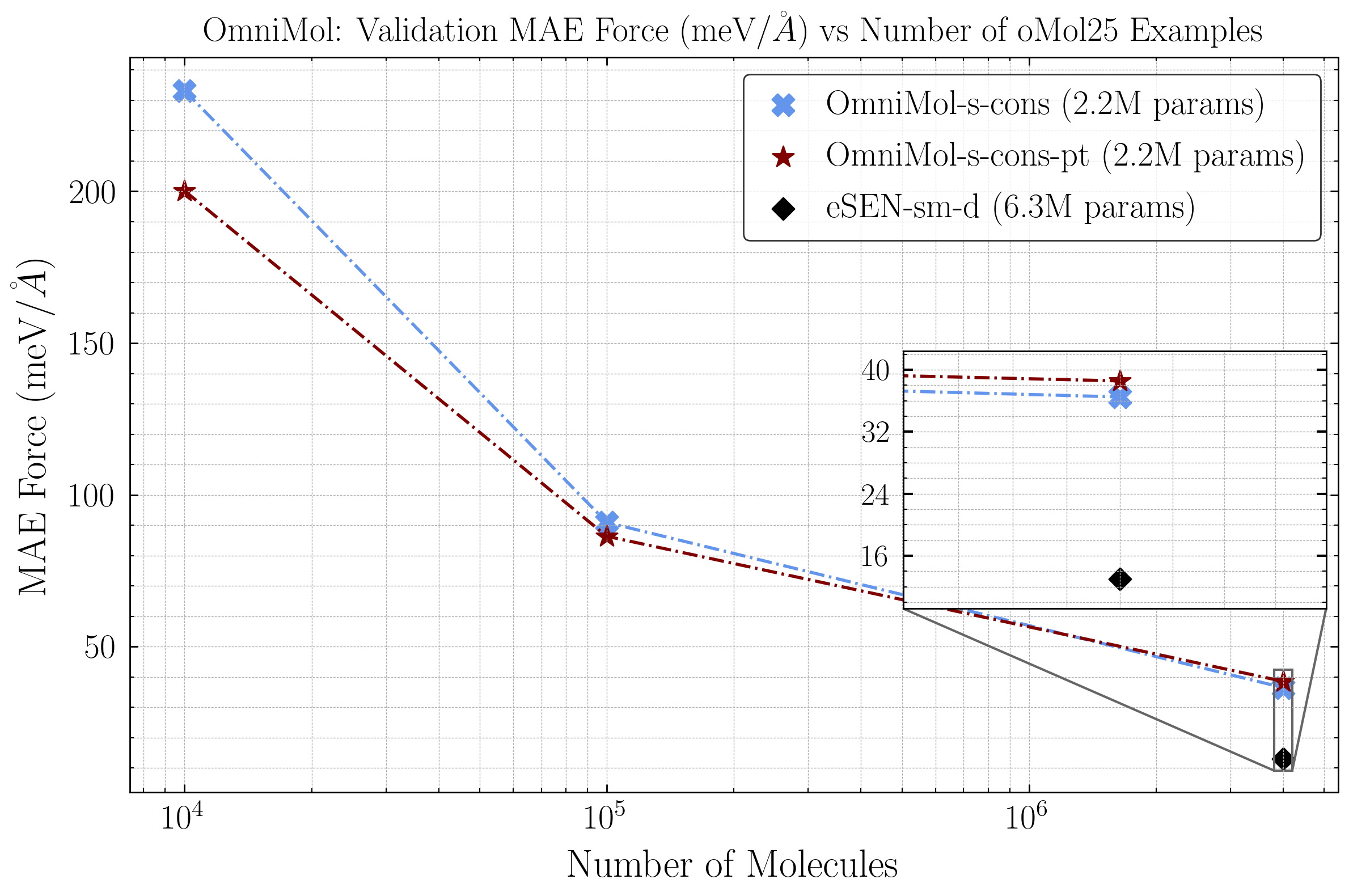}
    \caption{Scaling behavior for (left) energy and (right) forces of conservative and equivariant \textsc{OmniMol} small pre-trained and from scratch. Finetuning with ten and one hundred thousand molecules proceeds with LoRA, 4 million with full finetuning.}
    \label{fig:force_scaling}
\end{figure*}
\label{sec:results_fig}
We fine-tune a jet-pretrained PET backbone on oMoL using Low-Rank Adaptation (LoRA)~\cite{hu2021loralowrankadaptationlarge} applied \emph{only} to the transformer body. The base backbone consists of stacked transformer blocks with multi-head self-attention and feed-forward (MLP) sublayers. Each attention block involves learned projection matrices for queries, keys, values, and output (often denoted $W_Q, W_K, W_V, W_O$), while the MLP consists of linear layers with weights $W_{\mathrm{MLP}}$.

In our setup, all \emph{base} weights in the transformer body are frozen; we introduce LoRA adapters only on the body matrices, $W_Q, W_K, W_V, W_O, W_{\mathrm{MLP}}$. For each such matrix $W \in \mathbb{R}^{d_{\text{out}} \times d_{\text{in}}}$, we parameterize a low-rank update
\begin{equation}
W' = W + \Delta W, 
\qquad
\Delta W = \frac{\alpha}{r}\, B A,
\label{eq:lora-update}
\end{equation}
where $A \in \mathbb{R}^{r \times d_{\text{in}}}$ and $B \in \mathbb{R}^{d_{\text{out}} \times r}$ are trainable LoRA parameters, $r$ is the LoRA rank, and $\alpha$ is a scaling factor. In our experiments we set, $r = 96$ for all adapted matrices The original $W$ remain fixed with gradients disabled. All LoRA runs satisfy $\text{\#trainable parameters}  < 
\text{\#parameters from scratch}$.

Alongside the LoRA parameters, we train:
\begin{enumerate}
    \item the \emph{molecular encoders} that embed molecules into $\vec x_{embed} = \vec x_{embed}^{pos}+\vec x_{embed}^{Z} +\vec x_{embed}^{add}+\vec x_{embed}^{local}$.
    \item the \emph{bias MLP} that transforms all pairwise physics priors into a transformer bias $f(\vec r_i,\vec r_j,\vec x_{i,embed}^{Z},\vec x_{j,embed}^{Z})\rightarrow B_{ij}$.
    \item the \emph{task heads} that map the transformer representation to energy and force predictions.
\end{enumerate}
\paragraph{Embedding Adapters}
Finally, we introduce an "embedding adapting" layer. These are a per-token gated residual MLP placed in between the trained from scratch input encoders and the pre-trained transformers that modify learned embeddings $\vec x_{embed}$ by:
\begin{equation}
    \vec x_{embed}^* = \vec x_{embed} + \tanh(\alpha) * f_{MLP}(\vec x_{embed})\,.
\end{equation}
with a learnable gating scaler $\alpha$. We find this extra layer results in well-conditioned training. 
\subsection{Full Fine-Tuning}
We observe LoRA to be performant for finetuning on small subsets of oMol (>$\mathcal{O}(100k)$), but we find the benefits smaller at bigger architectures. Given that LoRA is generally best used for fast adaptation on small datasets this is expected \cite{hu2021loralowrankadaptationlarge,chen2022revisiting,ding2023delta}. 

To remedy this, we enable full finetuning. For full fine-tuning, we load \textit{all} weights that match in dimensionality in both the body and input encoders. We do not load any weights in the energy and force heads and we do not introduce embedding adapters. With full-finetuning, we find that the first epoch validation loss is significantly lower for pretrained models, approaching a factor of two improvement for \textsc{OmniMol} medium.

\section{Evaluation}
\label{sec:results}

In this section, we compare the performance of \textsc{OmniMol} to other MLIPs, comparing MAE of energies and forces.
\begin{table}[h]
    \centering
    \caption{Comparison between the performance reported for different algorithms trained on oMol 4M and evaluated on the Val-Comp Dataset. Bold results represent the algorithm with highest performance.}
    \label{tab:4m_subset_eval}
	\begin{tabular}{lccccc}
    \hline
          MAE $meV \downarrow$&  Energy \tiny{\textit{meV/Atom}} & Forces \tiny{$meV/\mathring{A}$}  \\
            \hline
            \textit{\textbf{GNN Baselines}} & & \\
            eSEN-sm-d (6.3M) & 2.19 & 13.01 \\
            eSEN-sm-cons (6.3M) & 1.89 & 11.10  \\
            eSEN-md-d (50.7 M) & \textbf{1.32}  & \textbf{6.78} \\
            \hline
            \textit{\textbf{All-to-All Transformers}} & & \\
            Transformer-1B (1B)\footnote{Model from \cite{kreiman2025transformersdiscovermolecularstructure}} & 1.99 & 18.35 \\
            TransIP (302M)\footnote{Model from \cite{elhag2025learninginteratomicpotentialsexplicit}} & 10.79 & 103.8
            \\ AllScAIP-sm-ft-cons (35 M)\footnote{Model from \cite{qu2026recipescalableattentionbasedmlips}}  & 1.04  & 8.19 \\
            AllScAIP-md-ft-cons (85 M)  & \textbf{0.90}  & \textbf{7.67 }\\
            \hline 
            \textit{\textbf{Ours}} & & \\
            \textsc{OmniMol-s-d} (2.2M) & 2.939 & 35.748   \\
            \textsc{OmniMol-s-d-pt} (2.2M) & 3.018 & 34.938   \\
            \textsc{OmniMol-s-cons} (2M) & 2.939 & 35.748   \\
            \textsc{OmniMol-s-cons-pt} (2M) & 3.29 & 38.55   \\
            \textsc{OmniMol-m-d} (43.3M) & \textbf{1.341} & \textbf{15.687}  \\
            \textsc{OmniMol-m-d-pt} (43.3M) & 1.441 & 16.980  
	\end{tabular}
\end{table}
\begin{table}[h]
    \centering
    \caption{Comparison between the performance reported for different algorithms trained on oMol 100M/140M and evaluated on the Val-Comp Dataset. Bold results represent the algorithm with the highest performance.}
    \label{tab:100m_eval}
	\begin{tabular}{lccccc}
    \hline
          MAE $meV \downarrow$&  Energy \tiny{\textit{meV/Atom}} & Forces \tiny{$meV/\mathring{A}$}  \\
            \hline
            \textit{\textbf{Baselines}} & & \\
            eSEN-sm-d (6.3M) & 1.49 & 9.92 \\
            eSEN-md-d (50.7 M) & 0.84  & \textbf{4.76} \\
            AllScAIP-md-d (85 M)\footnote{Model from \cite{qu2026recipescalableattentionbasedmlips}}  & \textbf{0.64}  & 5.24 \\
            \hline
            \textit{\textbf{Ours}} & & \\
            \textsc{OmniMol-m-d} (43.3M) & 1.263 & 14.47  \\
            \textsc{OmniMol-l-d} (306.3M) & \textbf{1.04} & \textbf{13.59 } 
	\end{tabular}
\end{table}In Table \ref{tab:4m_subset_eval} and Table \ref{tab:100m_eval}, we compare the performance of \textsc{OmniMol-s/m/l} to other MLIPs when fully trained on oMol-4M and oMol-100M/140M, where we find strong performance. In Table \ref{tab:100_subset_eval} and Table \ref{tab:2_epoch_eval}, we observe the significant benefit of jet-physics pre-training when training with a small number of molecules ($\sim100k$ or less) or with a small number of passes over oMol-4M (two epochs in our example). In this low-compute or low-data regime, pre-training can improve performance by up to 50\%, potentially reducing the required compute for a given performance target. The pre-training advantage is reduced as more molecular training examples and more compute is provided as seen in Table \ref{tab:4m_subset_eval}. We also observe the benefit of the equivariant + conservative model, with significantly improved performance  (especially forces) when training with the small subset of molecules. We again observe this raw MAE advantage of the equivariant + conservative model disappearing with large number of molecules. This provides valuable intuition for how to understand jet pre-training: a strong inductive bias, and like other inductive bias (such as equivariance/invariance), most useful when training data is limited, i.e. the bitter lesson \cite{sutton2019bitterlesson,yousefi2024learningbitterlesson,wu2025bitterlessonmultilingual}. 

\begin{table}[h]
    \centering
    \caption{Comparison between the performance of 
    \textsc{OmniMol} variants trained on a \textbf{100k subset} of oMol and evaluated on the Val-Comp Dataset. Bold results represent the algorithm with highest performance.}
    \label{tab:100_subset_eval}
	\begin{tabular}{lcc}
    \hline
          MAE $meV \downarrow$&  Energy \tiny{\textit{meV/Atom}} & Forces \tiny{$meV/\mathring{A}$}  \\
    \hline
    \textsc{OmniMol-s-d} (2.2M) & 13.598 & 122.918 \\
    \quad\scriptsize\textit{Pre-Training Advantage (\%)} & \scriptsize\textit{+12.3\%} & \scriptsize\textit{+19.5\%} \\
    \textsc{OmniMol-s-d-pt} (2.2M) & \textbf{11.924} & \textbf{98.995} \\
    \hline
    \textsc{OmniMol-s-cons} (2M) & 12.338 & 90.925 \\
    \quad\scriptsize\textit{Pre-Training Advantage (\%)} & \scriptsize\textit{+6.1\%} & \scriptsize\textit{+5.0\%} \\
    \textsc{OmniMol-s-cons-pt} (2M) & \textbf{11.586} & \textbf{86.363} \\
    \hline
    \textsc{OmniMol-m-d} (43.3M) & 11.582 & 99.758 \\
    \quad\scriptsize\textit{Pre-Training Advantage (\%)} & \scriptsize\textit{+29.4\%} & \scriptsize\textit{+26.9\%} \\
    \textsc{OmniMol-m-d-pt} (43.3M) & \textbf{8.175} & \textbf{72.876} \\
    \hline
	\end{tabular}
\end{table}
\begin{table}[h]
    \centering
    \caption{Comparison between the performance of 
    \textsc{OmniMol} variants trained with \textbf{only two passes} of oMol 4M and evaluated on the Val-Comp Dataset. Bold results represent the algorithm with highest performance.}
    \label{tab:2_epoch_eval}
	\begin{tabular}{lcc}
    \hline
          MAE $meV \downarrow$&  Energy \tiny{\textit{meV/Atom}} & Forces \tiny{$meV/\mathring{A}$}  \\
    \hline
    \textsc{OmniMol-s-d} (2.2M) & 73.767 & 395.76 \\
    \quad\scriptsize\textit{Pre-Training Advantage (\%)} & \scriptsize\textit{+34.7\%} & \scriptsize\textit{+29.4\%} \\
    \textsc{OmniMol-s-d-pt} (2.2M) & \textbf{48.174} & \textbf{279.29} \\
    \hline
    \textsc{OmniMol-s-cons} (2M) & \textbf{84.62} & 252.77 \\
    \quad\scriptsize\textit{Pre-Training Advantage (\%)} & \scriptsize\textit{-20.1\%} & \scriptsize\textit{+17.6\%} \\
    \textsc{OmniMol-s-cons-pt} (2M) & 101.59 & \textbf{208.34} \\
    \hline
    \textsc{OmniMol-m-d} (43.3M) & 62.34 & 298.63 \\
    \quad\scriptsize\textit{Pre-Training Advantage (\%)} & \scriptsize\textit{+54.6\%} & \scriptsize\textit{+56.9\%} \\
    \textsc{OmniMol-m-d-pt} (43.3M) & \textbf{28.31} & \textbf{128.57} \\
    \hline
	\end{tabular}
\end{table}

\subsection{All-to-All Transformer Benefits: Scaling and Compute}
\begin{figure*}[t]
    \centering
    \begin{subfigure}[t]{0.49\textwidth}
        \centering
        \includegraphics[width=\textwidth]{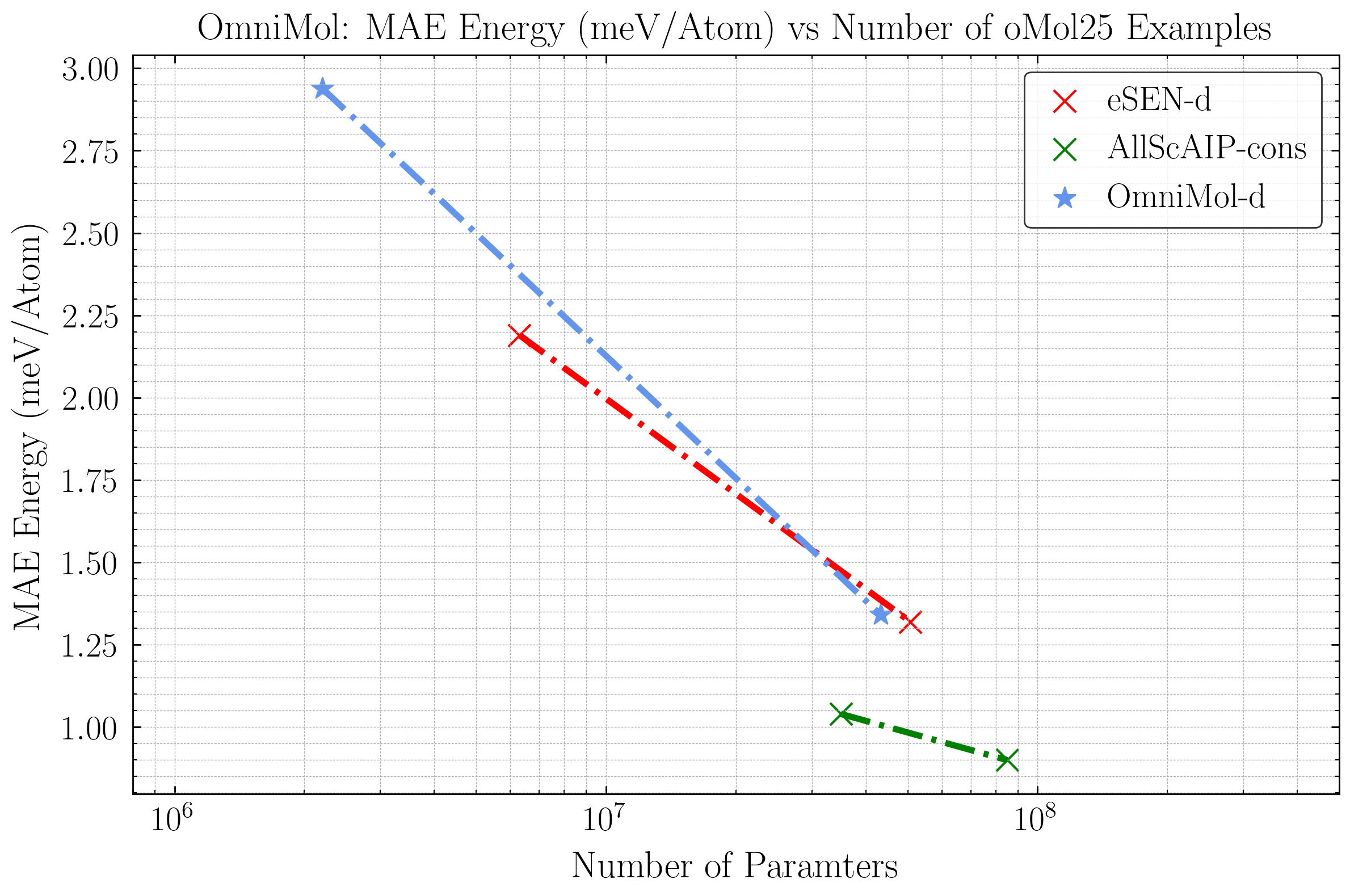}
        \caption{Scaling behavior for energy with respect to model size for \textsc{OmniMol}-d, eSEN-d, and AllScAIP-md-cons, trained on oMol-4M}
        \label{fig:model_scaling_energy}
    \end{subfigure}
    \hfill
    \begin{subfigure}[t]{0.49\textwidth}
        \centering
        \includegraphics[width=\textwidth]{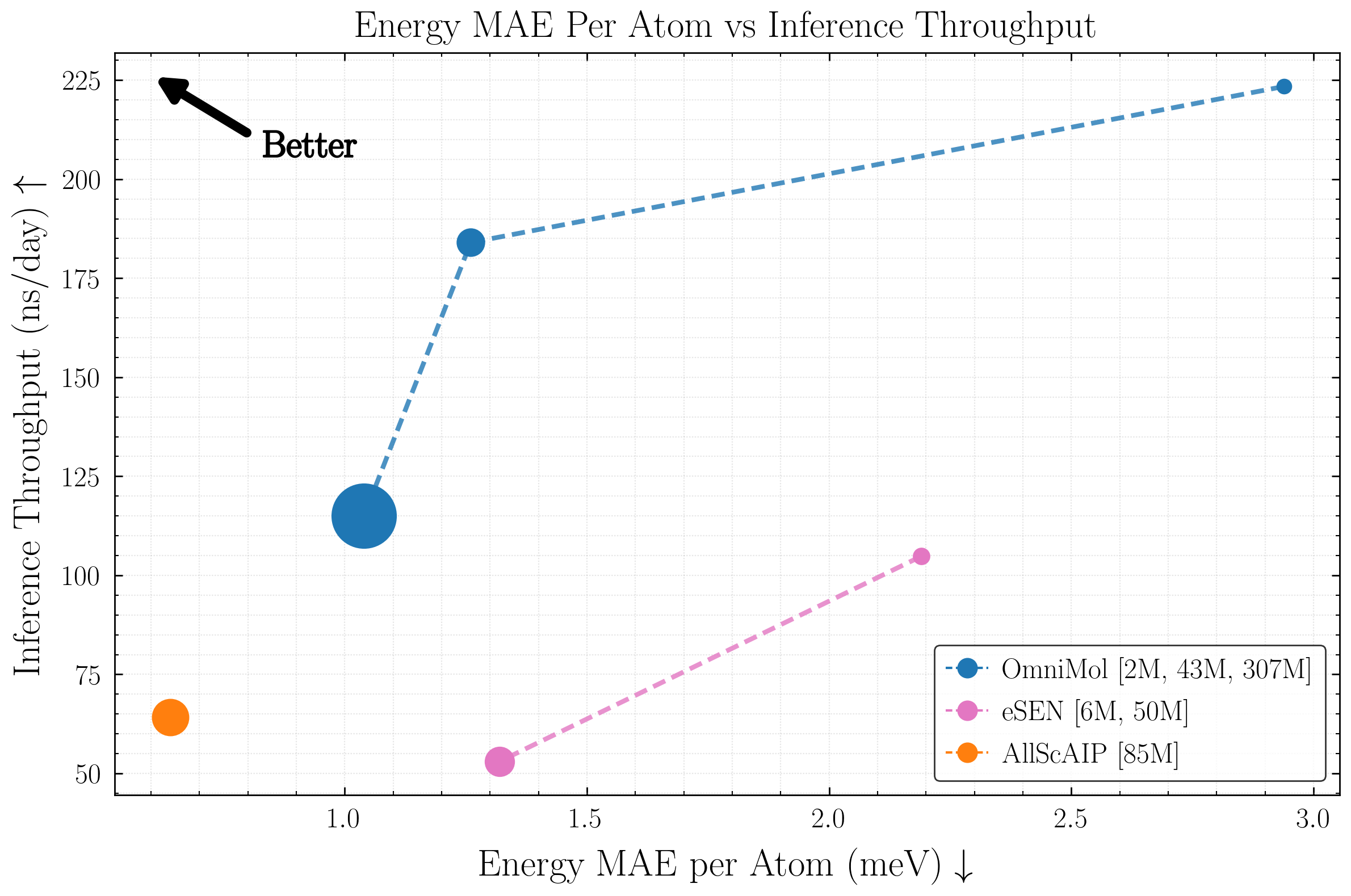}
        \caption{Inference Speed–Energy Error Plane. All models evaluated on $\mathcal{O}(100)$ atom systems on one A100-40GB GPU}
        \label{fig:inference_speed}
    \end{subfigure}
    \caption{Scaling behavior for energy with respect to model size for \textsc{OmniMol}-d, eSEN-d, and  AllScAIP-md-cons (left), and inference speed–energy error trade-off for \textsc{OmniMol}-d, eSEN-d, and AllScAIP-md-d (right).}
    \label{fig:model_scaling}
\end{figure*}
\label{sec:conc}
We would like to highlight the scaling and inference speed of \textsc{OmniMol}. Transformers are known to be data and parameter-hungry and often under perform relative to other architectures if either is insufficient \cite{dosovitskiy2020vit, touvron2021deit}. On the other hand, they seem to be superior in scaling, outperforming architectures with better priors with sufficient data and size \cite{kaplan2020scalinglaws,hoffmann2022chinchilla,zhai2022scalingvit,kreiman2025transformersdiscovermolecularstructure,qu2026recipescalableattentionbasedmlips}. This pattern appears to extend to MLIPs as well. Recent work shows that transformers for molecular energies and forces follow clean power-law scaling with model size and show no signs of saturation up to 1B parameters, and thus their scaling fits can successfully extrapolate to larger models \cite{kreiman2025transformersdiscovermolecularstructure,qu2026recipescalableattentionbasedmlips}. We show model size scaling in in Figure \ref{fig:model_scaling_energy}.


Importantly, the same studies also finds that these transformers can train and evaluate faster (in wall-clock time) than sparse message-passing GNNs despite large parameter counts, benefiting from the plethora of hardware optimizations built for transformers \cite{kreiman2025transformersdiscovermolecularstructure,qu2026recipescalableattentionbasedmlips}. To test this explicitly, we compare the inference speed of GNNs and transformers on one A100-40GB GPU on systems of $\mathcal{O}(100)$ atoms. The results are shown in Table \ref{tab:speed_eval} and Figure \ref{fig:inference_speed}. \textsc{OmniMol} is unique on the inference speed-error plane, not the most accurate, but, for the case of \textsc{OmniMol-m}, $\sim$3x faster than eSEN-md-d and AllScAIP-md, while only introducing a $\sim.7$ energy MAE/atom error delta from AllScAIP-md (and while outperforming eSEN-md-d). This is non-insignificant. In applications such as small molecule drug searches, rapid exploration of the design space is required. A 3x inference speed thus translates to exploring 3x more molecules on the same compute budget. 
\vspace{-0em}
\begin{table}[h]
    \centering
    \caption{Comparison between inference speed for different algorithms trained on oMol 4M/100M/140M and evaluated on a $\mathcal{O}(100)$ atom systems and one A100-40GB GPU. Atom $\cdot~ ns$/day is computed with an implicit MD timescale of $1$ femtosecond per forward pass. Bold results represent the algorithm with the highest performance.}
    \label{tab:speed_eval}
	\begin{tabular}{lccccc}
    \hline
          Inference Speed &  Forward $ms \downarrow$ &~ atom $\cdot~ ns$/day  $\uparrow$\\
            \hline
            \textit{\textbf{Baselines}} & & \\
            eSEN-sm-d (6.3M) & \textbf{17.314} & \textbf{104.80} \\
            eSEN-md-d (50.7 M) & 34.199 & 53.05 \\
            AllScAIP-md-d (85 M)\footnote{Model from \cite{qu2026recipescalableattentionbasedmlips}}  & 28.289  & 64.14 \\
            \hline
            \textit{\textbf{Ours}} & & \\
            \textsc{OmniMol-s-d} (43.3M) & 	\textbf{8.120} & \textbf{223.40} \\
            \textsc{OmniMol-m-d} (43.3M) & 9.861 & 184.00 \\
            \textsc{OmniMol-l-d} (306.3M) & 15.764 & 115.10
	\end{tabular}
\end{table}

\section{Conclusion and Outlook}
\label{sec:conclusions}
We present \textsc{OmniMol}, a cross-domain machine-learned interatomic potential obtained by adapting a particle physics jet-pretrained Point-Edge Transformer (PET) backbone to molecular energies and forces. We studied both parameter-efficient adaptation (LoRA) and full fine-tuning from jet physics pre-training, showing that cross-domain initialization can substantially accelerate early optimization and improve accuracy in the low-data regime. We further demonstrate that due to architectural transfer from \textsc{OmniLearned}, we can achieve uniquely fast inference speeds with potentially sufficiently small error rate.
Looking ahead, we leave several important directions to future work. We plan to broaden evaluation beyond static energy and force error to better characterize where cross-domain transfer most reliably improves scientific utility.

Even though we focus on MLIPs, the lessons learned here could have widespread utility.  Many scientific domains describe the systems of interest as unordered sets of interacting bodies, each endowed with attributes. Point clouds are widespread in science and it would be interesting to explore further connections in the future.
\vspace{-1.5em}
\section*{Code Availability}

The code for this paper can be found at \url{https://github.com/ibrahimEls/OmniMol}.

\section*{Acknowledgments}
We thank Samuel Blau, Ray Gao, Brandon Wood, Daniel Levine, and Abdulrahman Aldoosary for interesting discussions which significantly deepened our knowledge of the needs of the field. In addition, we thank Joschka Birk for many lessons in model development. VM is supported by JST EXPERT-J, Japan Grant Number JPMJEX2509.
BN is supported by the U.S. Department of Energy (DOE), Office of Science under contract DE-AC02-76SF00515.  This research used resources of the National Energy Research Scientific Computing Center, a DOE Office of Science User Facility supported by the Office of Science of the U.S. Department of Energy under Contract No. DE-AC02-05CH11231 using NERSC awards ERCAP0034229, HEP-ERCAP0021099 and HEP-ERCAP0028249.

\newpage
\bibliography{bib}
\bibliographystyle{apsrev4-1}

\appendix

\clearpage

\end{document}